\documentclass[5p,,preprint,12pt]{elsarticle}
\makeatletter\if@twocolumn\PassOptionsToPackage{switch}{lineno}\else\fi\makeatother

\usepackage{tabulary,xcolor}
\usepackage{svg}
\usepackage{amsfonts,amsmath,amssymb}
\usepackage[T1]{fontenc}
\usepackage{gensymb}
\usepackage{siunitx}
\usepackage{dblfloatfix}
\usepackage[none]{hyphenat} %Gets rid of putting words on multiple lines from hyphenation.
\usepackage{ragged2e}
\usepackage{setspace} 

\makeatletter
\let\save@ps@pprintTitle\ps@pprintTitle
\def\ps@pprintTitle{\save@ps@pprintTitle\gdef\@oddfoot{\footnotesize\itshape \null\hfill\today}}
\def\hlinewd#1{%
  \noalign{\ifnum0=`}\fi\hrule \@height #1%
  \futurelet\reserved@a\@xhline}

\AtBeginDocument{\ifNAT@numbers \biboptions{sort&compress}\fi}

\makeatother

\usepackage{ifluatex}
\ifluatex
\usepackage{fontspec}
\defaultfontfeatures{Ligatures=TeX}
\usepackage[]{unicode-math}
\unimathsetup{math-style=TeX}
\else 
\usepackage[utf8]{inputenc}
\fi 
\ifluatex\else\usepackage{stmaryrd}\fi

\usepackage{url,multirow,morefloats,floatflt,cancel,tfrupee}
\makeatletter

\AtBeginDocument{\@ifpackageloaded{textcomp}{}{\usepackage{textcomp}}}
\makeatother
\usepackage{todonotes}
\usepackage{colortbl}
\usepackage{xcolor}
\usepackage{pifont}
\usepackage[nointegrals]{wasysym}
\makeatletter

\def\mcWidth#1{\csname TY@F#1\endcsname+\tabcolsep}

\def\cAlignHack{\rightskip\@flushglue\leftskip\@flushglue\parindent\z@\parfillskip\z@skip}
\def\rAlignHack{\rightskip\z@skip\leftskip\@flushglue \parindent\z@\parfillskip\z@skip}

\@ifundefined{etal}{}{}

\usepackage{ifxetex}
\ifxetex\else\if@twocolumn\@ifpackageloaded{stfloats}{}{\usepackage{dblfloatfix}}\fi\fi

\AtBeginDocument{
\expandafter\ifx\csname eqalign\endcsname\relax
\def\eqalign#1{\null\vcenter{\def\\{\cr}\openup\jot\m@th
  \ialign{\strut$\displaystyle{##}$\hfil&$\displaystyle{{}##}$\hfil
      \crcr#1\crcr}}\,}
\fi
}

\AtBeginDocument{%
  \@ifpackageloaded{endfloat}%
   {\renewcommand\efloat@iwrite[1]{\immediate\expandafter\protected@write\csname efloat@post#1\endcsname{}}}{\newif\ifefloat@tables}%
}%

\def\BreakURLText#1{\@tfor\brk@tempa:=#1\do{\brk@tempa\hskip0pt}}
\let\lt=<
\let\gt=>
\def\processVert{\ifmmode|\else\textbar\fi}

\@ifundefined{subparagraph}{
\def\subparagraph{\@startsection{paragraph}{5}{2\parindent}{0ex plus 0.1ex minus 0.1ex}%
{0ex}{\normalfont\small\itshape}}%
}{}

\newcommand\role[1]{\unskip}
\newcommand\aucollab[1]{\unskip}
  
\@ifundefined{tsGraphicsScaleX}{\gdef\tsGraphicsScaleX{1}}{}
\@ifundefined{tsGraphicsScaleY}{\gdef\tsGraphicsScaleY{.9}}{}
\def\checkGraphicsWidth{\ifdim\Gin@nat@width>\linewidth
	\tsGraphicsScaleX\linewidth\else\Gin@nat@width\fi}

\def\checkGraphicsHeight{\ifdim\Gin@nat@height>.9\textheight
	\tsGraphicsScaleY\textheight\else\Gin@nat@height\fi}

\def\fixFloatSize#1{}%\@ifundefined{processdelayedfloats}{\setbox0=\hbox{\includegraphics{#1}}\ifnum\wd0<\columnwidth\relax\renewenvironment{figure*}{\begin{figure}}{\end{figure}}\fi}{}}
\let\ts@includegraphics\includegraphics

\def\inlinegraphic[#1]#2{{\edef\@tempa{#1}\edef\baseline@shift{\ifx\@tempa\@empty0\else#1\fi}\edef\tempZ{\the\numexpr(\numexpr(\baseline@shift*\f@size/100))}\protect\raisebox{\tempZ pt}{\ts@includegraphics{#2}}}}

\AtBeginDocument{\def\includegraphics{\@ifnextchar[{\ts@includegraphics}{\ts@includegraphics[width=\checkGraphicsWidth,height=\checkGraphicsHeight,keepaspectratio]}}}

\DeclareMathAlphabet{\mathpzc}{OT1}{pzc}{m}{it}

\def\URL#1#2{\@ifundefined{href}{#2}{\href{#1}{#2}}}

\def\UrlOrds{\do\*\do\-\do\~\do\'\do\"\do\-}%
\g@addto@macro{\UrlBreaks}{\UrlOrds}

\edef\fntEncoding{\f@encoding}

\makeatother

\newif\ifmultipleabstract\multipleabstractfalse%
    \makeatletter
\def\ead{\@ifnextchar[{\@uad}{\@ead}}
\gdef\@ead#1{\bgroup
   \def\_{\string\underscorechar\space}
   \def\{{\string\lbracechar\space}
   \def\textdagger{\string\textdagger\space}
   \def\texttildeapprox{\string\texttildeapprox\space}
   \def~{\hashchar\space}
   \def\}{\string\rbracechar\space}
   \edef\tmp{\the\@eadauthor}
   \immediate\write\@auxout{\string\emailauthor
     {#1}{\expandafter\strip@prefix\meaning\tmp}}
  \egroup
}
\gdef\emailauthor#1#2{\stepcounter{ead}
      \g@addto@macro\@elseads{\justifying
      \let\corref\@gobble
      \eadsep\texttt{#1} (#2)
      \def\eadsep{\unskip,\space}}
}

\makeatother

\begin{document}
\begin{frontmatter}

\title{Microstructural characterization to reveal evidence of shock deformation in a Campo del Cielo meteorite fragment}
\author{Graeme J. Francolini\corref{correspondingauthor}}
\cortext[correspondingauthor]{Corresponding author.}
\ead{graemejf@mail.ubc.ca}
\author{Thomas B. Britton}
\ead{ben.britton@ubc.ca}
\address{Department of Materials Engineering, University of British Columbia, BC, Canada}
\date{\today}                     %% if you don't need date to appear

\begin{abstract}
\par For materials scientists and engineers, the extreme and unusual conditions which meteorites and their microstructures form allow for insight into materials which would exist at the edge of our thermomechanical processing abilities. One such microstructure found in low-shock event iron meteorites is Neumann bands. These bands are arrays of lenticular deformation twins that form throughout the Fe-Ni matrix with numerous intersections, resulting in many high stress and strain regions within the material's surface. These regions and the shocks that formed them encourage atypical strain accommodating mechanisms and structural changes of the material. However, investigation of the deformation twin intersections and the microstructural behaviour in and around these regions has been limited. In this work, investigation of these regions in a Campo del Cielo meteorite fragment, with electron backscatter diffraction (EBSD) and forescatter electron (FSE) imaging, revealed two primary findings: high-intensity pattern doubling mirrored across the \{110\} band at twin-twin intersection and microband formation across the sample surface, suggesting multilayer twinning and constraint of the crystal structure at twin-twin intersection points. Microbands were found to form along the \{110\} plane and in regions near Neumann bands. The simultaneous existence of Neumann bands (microtwins) and microbands is presented here for a BCC material, and it is believed the Neumann band and microbands formed during different types and/or shock events from one another. The presence of both Neumann bands and microbands within a BCC iron meteorite is previously unreported and may be valuable in furthering our understanding of shock deformation within iron-based materials.
\end{abstract}

  \end{frontmatter}

\onecolumn   %TBBEDIT_PROOF  
\section{Introduction}
\par Meteorites and their fragments offer valuable insights into the processes governing material formation both on Earth and in extraterrestrial environments. In brief, a meteorite is a meteor that survives its passage through the atmosphere of the earth and strikes the ground. These meteorites often contain very interesting microstructures that are uncommonly found on earth, as they have unique chemistry and thermomechanical processing histories (e.g. very long cooling periods) that depend on the meteor's formation, movement through space, and location of impact. When the meteor becomes a meteorite, it may experience short-lived rapid shock which can result in some unusual characteristic microstructural features as deformation occurs at very high strain rates (estimated between \SI{e2}{\per\second} and \SI{e8}{\per\second}) \cite{Ramesh2008}, often with internal aftershocks. Therefore, the ultimate microstructure of a meteorite fragment is dependent on the impact and thermal histories of the host meteor and the resulting meteorite \cite{Scott1981, Luo2024}, in a similar manner to how the microstructure of an engineering component is influenced by the manufacturing process. These high strain rate  microstructures, resulting from impact shocks, can be characterized to understand these otherworldly materials, and provide insight into the extremes of our accessible thermomechanical processing envelope.

\par The present manuscript focuses on the Campo del Cielo meteorite, which is an iron-rich Fe-Ni metallic meteorite. Major phases in the Fe-Ni system include kamacite ($\alpha$) and taenite ($\gamma$). The kamacite phase is closely related to the ferrite phase in steel, with a body centred cubic (BCC) crystal structure and a chemistry that is Fe-rich with typically 3.3-\SI{5.8}{wt\%Ni} \cite{Nichiporuk1958, Perry1944}. The taenite phase is closely related to the austenite phase in steel, with a face centered cubic (FCC) crystal structure and a chemistry that is Fe-rich with typically 31 to \SI{54}{wt\%Ni} \cite{Nichiporuk1958, Perry1944}. Additionally, the presence of other elements, such as P, S, and Co, results in the precipitation of different minerals within the iron matrix. A common mineral precipitate that is found is schreibersite (also known as rhabdite), an iron-nickel phosphide mineral with a tetragonal crystal structure and appears as a blocky precipitate. This precipitate is found to form in preferential orientations, with the orientation relationship given in Eqn. \ref{Eqn:ORPlane} and \ref{Eqn:ORDirection} \cite{Hennig1999}. Iron meteorites are also known to have other microstructural features which form in preferential orientations, such as Neumann bands.

\begin{equation}
\label{Eqn:ORPlane}
\{110\} (Fe,Ni)_3 P || \{210\} \alpha-(Fe,Ni)
\end{equation}
\begin{equation}
\label{Eqn:ORDirection}
<001> (Fe,Ni)_3 P || <001> \alpha-(Fe,Ni)
\end{equation}

\par Neumann bands are an observed form of deformation twinning, where the twins are found within the kamacite phase along the \{112\} planes. The formation of these bands is a direct result of the shock incurred from the meteorite's impact \cite{Luo2024}, with each twin providing a high rate strain accommodation mechanism that will be formed as one of twelve possible twin variants with respect to the parent BCC-grain \cite{Smith1928, Uhlig1955}. For these high rate deformation twins in BCC-kamacite, it is thought that the nuclei of these twins first form due to an accumulation of stacking faults along \{112\} planes \cite{Groger2023}, with twin propagation initiating once a critical shear stress value is reached \cite{Koko2021}. The twins are thought to form as an oblate spheroid, which present within a 2D planar section as elongated ovoids.
%\begin{figure}[!hb]
%    \centering
%    \includegraphics[width=0.5\textwidth]{Advanced Figures/NeumannBandIllustration.png}
%    \caption{Drawn illustration of Neumann bands overlapping at various orientations. These bands form along the \{112\} planes of BCC materials through shock events and shear deformation. }
%    \label{Fig:NeumannBandIllustration}
%\end{figure}{}
\par Neumann bands tend to extend (at least) across the surface of iron meteorites and exist in various thicknesses, lengths and crystal orientations. The twins tend to form in groups of similar direction and orientation, but they do not all form simultaneously. Rather, families of Neumann bands form as a result of different shock waves that move through the meteorite and it is likely that each band that extends in the same direction within a large single crystal matrix is related to the individual variant of the deformation twin. When multiple families of bands form, this results in twin growth and intersections through one another.  

\par The growth and intersection of these bands in kamacite, and other BCC metals, are dependent on the twinning, slip and shear planes within the metal-matrix grain. After the initial formation of a band, it will undergo forward growth along the twin shear direction, \{112\}, through deformation twinning. As the twin grows it generates highly strained regions within the matrix, thereby raising the driving stress needed for further propagation \cite{Zhong2024, Britton2015, Wang2022}. This increase in required driving stress leads to dislocation mobility and slip being a more energetically favourable deformation mechanism. Resulting in outward growth and thickening of the Neumann bands until forward growth becomes energetically favourable again.  

\par Single slip deformation in BCC metals is also known to result in the formation of another visually unique microstructural feature called microbands \cite{Thuillier1994, Huang1989, Jackson1983}. Microbands appear as linear bands, coincident with the \{110\} planes in BCC metals \cite{Pappu2000}, and form initially from strain-induced primary slip planes with independent dislocations. Shear stress acting on these slip planes then results in a polarized arrangement of the dislocations and the formation of double dislocation walls. A more in-depth explanation of the microband formation mechanism is given by \emph{Huang and Gray} \cite{Huang1989}. While these bands are mechanistically independent from Neumann bands, they may share traces of \{110\} with Neumann band traces of \{112\} along the [110] orientation \cite{Pappu2000}. With that said, microbands and Neumann bands have been previously observed within iron and steel materials separately \cite{Shen2007, Gutierrez-Urrutia2022} and in lab-based impact tests in stainless steel "crater-related microstructures are dominated by deformation microtwins with a few intermixed microbands" \cite{Murr2002}. However, we are not aware of previously documented simultaneous existence of these microstructural features for iron meteorites. 

\par The Neumann bands, their twelve possible formation directions and their frequent intergrowth result in a complex microstructure which contains twin intersections with high strain gradients. These intersections and the regions surrounding them are of high interest for investigation using Electron Backscatter Diffraction (EBSD). Previous works have used this technique to investigate iron meteorites \cite{Luo2024, Cayron2014, Nolze2005}. However, there has been limited study of the intersections of the deformation twins and the microstructural behaviour in these regions. 

\par The present work aims to remedy this by further investigating the high stress and strain areas created by the Neumann bands and their intersections present within the microstructure of iron-based meteorites. In particular, abnormal material behaviour and phenomena are to be investigated using EBSD. Similar works have been performed previously for both titanium deformation twins \cite{Guo2017} and for other iron-based materials \cite{Koko2021}. However, there is a lack of investigation into these areas for the Campo del Cielo meteorite group.

\section{Materials and Methods}

\subsection{Material Preparation}

\par A meteorite fragment from the Campo del Cielo meteorite group (sourced from Rocks $\&$ Gems Canada, Whistler, BC) was metallographically polished using 400, 600, 800, and 1200 grit SiC for 10 minutes each, with the sample rotated \SI{90}{\degree} after 5 minutes per grit. This was followed by polishing with a \SI{6}{\micro m} and \SI{1}{\micro m} diamond suspension, and a \SI{0.05}{\micro m} colloidal silica suspension for \SI{6}{minutes} each (again rotating by \SI{90}{\degree} halfway through). Lastly, broad ion beam (Ar-ions) polishing was performed using a Gatan PECS II system for 1 hour at \SI{4}{\degree} tilt, \SI{2}{keV}, \SI{1}{RPM}, and no modulation.

\subsection{Material Analysis}

\par Electron microscopy was performed on a Tescan AMBER-X plasma focused ion beam-scanning electron microscope (pFIB-SEM). Secondary Electron (SE) images were obtained using a \SI{10}{nA} beam current, \SI{20}{kV} accelerating voltage, and a working distance of 13-14 \SI{}{mm}. EBSD was performed with an Oxford Instruments Symmetry S2 detector using \SI{10}{nA} beam current and \SI{20}{kV} accelerating voltage, with the sample loaded into a \SI{70}{\degree} pre-tilted holder. EBSD acquisition was performed three times with increased magnification on regions of interest (ROI). The first ROI was imaged with an acquisition speed of \SI{1613.6}{Hz}, a resolution of 156x\SI{128}{pixels}, a step size of \SI{0.15}{\micro m} and an exposure time of \SI{0.60}{ms}, the second ROI at \SI{164.5}{Hz}, 1244x\SI{1024}{pixels}, \SI{50}{nm} and \SI{6.00}{ms}, and the third ROI at \SI{91.3}{Hz}, 622x\SI{512}{pixels}, \SI{50}{nm} and \SI{8.88}{ms}. Further Forescatter electron (FSE) imaging was performed with an increased dwell time of \SI{65}{\micro s} at a working distance of \SI{14.0}{mm}, \SI{20}{kV}, and \SI{10}{nA}.
\par Post-acquisition analysis of the EBSD data was performed using Aztec Crystal, using MapSweeper which performs a 'template matching' approach of comparing each captured diffraction pattern with a dynamical-diffraction based simulation and this improves the indexing precision \cite{Winkelmann2020}, and subsequent analysis within MTEX 5.10.2 \cite{MTEX} in MATLAB. For one map (shown in Figure 3), high (angular) resolution EBSD (HR-EBSD) analysis was performed using CrossCourt Rapide v4.6 to perform total GND analysis in the BCC-phase assuming <110> dislocation types, using remapping based cross correlation with 40 ROIs and \cite{Britton2012}. 

\par The sample was mounted flat for energy dispersive X-ray spectroscopy (EDS/EDX), and area, point mapping was performed using an Oxford Instruments EDS analysis using an Ultim Max 170 detector. Analysis was performed at \SI{20}{kV}, \SI{1}{nA}, and a working distance of \SI{8.0}{mm}, using Oxford Instruments Aztec with the Pulse Pile up correction. 

\par Simulated EBSP were created through reprojection of dynamical pattern simulations. DynamicS was used to generate a reference many beam dynamical EBSD pattern simulation, using a BCC-iron crystal structure (a = \SI{0.286}{nm}). These simulations were based upon the work of Winkelmann \emph{et al.} \cite{Winkelmann2007} and included reflectors with a lower limit of d\textsubscript{hkl} = \SI{0.03}{nm}, or the intensity of the bands being \SI{20}{\percent} of the strongest reflector. The reference sphere was imported into AstroEBSD \cite{AstroEBSD, Britton2018b} within MATLAB and the experimental pattern was reprojected using the same pattern centre and crystal orientation as in Fig. \ref{Fig:SmallROI} c).

\section{Results}

\par
\begin{figure*}[!htb]
    \centering
    \includegraphics[width=\textwidth]{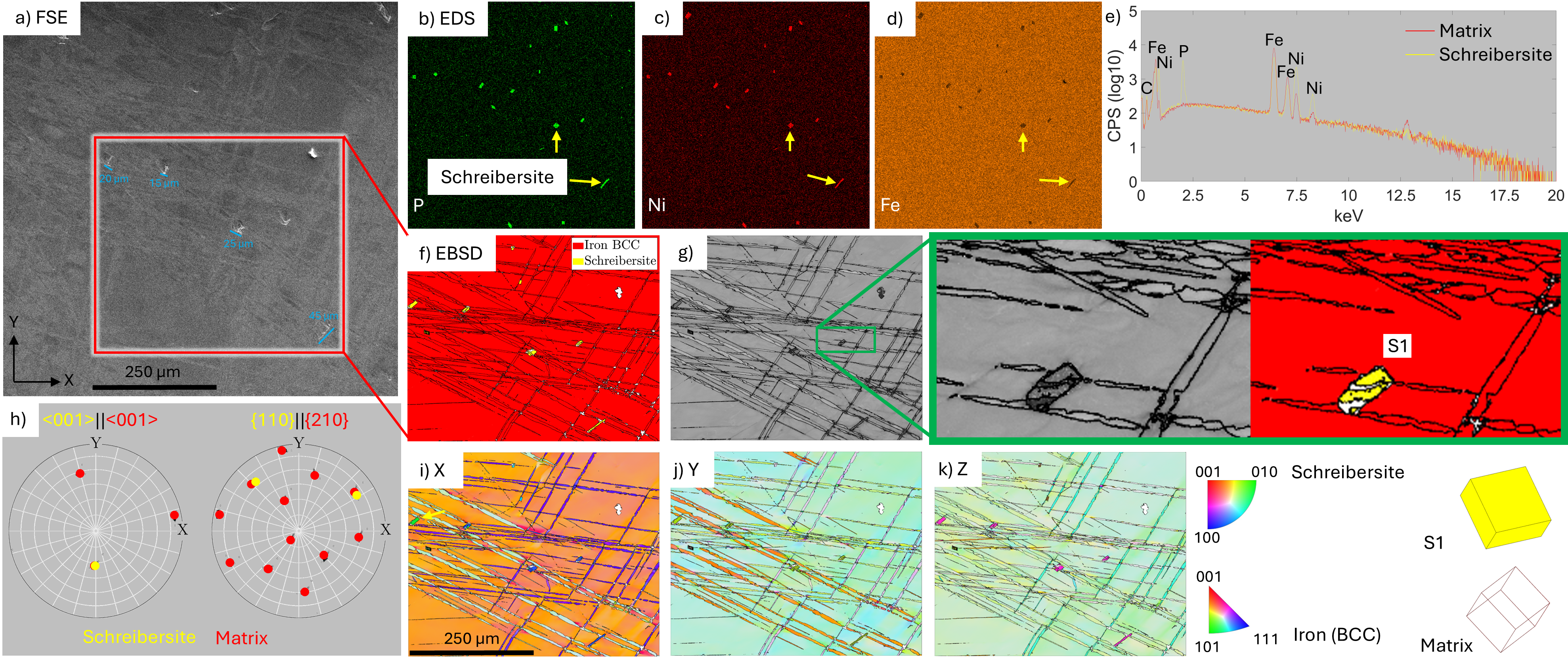}
    \caption{a) Lower magnification scanning electron microscopy (SEM), Electron Backscatter Diffraction (EBSD) and Energy-dispersive X-ray spectroscopy (EDS) of Campo del Cielo fragment. A large amount of intersection is seen in the Neumann bands and the schreibersite precipitates are seen to form in preferential orientations along the bands. EDS confirms the surface chemistry is as expected for iron-meteorites. First row: EDS phase maps (b-d) and e) EDS spectra for matrix and schreibersite. Second row: f) phase map and g) band contrast. Third row: h) Overlaid pole figures of schreibersite and kamacite matrix, inverse Pole Figures (IPF) (i-k), and IPF keys and unit cells for schreibersite and iron (BCC). The overlaid pole figures are schreibersite <001> direction with matrix <001>, and schreibersite \{110\} with matrix \{210\} pole figures.}
    \label{Fig:BigROI}
\end{figure*}

 Fig. \ref{Fig:BigROI} shows an overview of the area studied, and reveals that the meteorite fragment was found to be composed of kamacite with a large number of cuboid schreibersite precipitates, as confirmed with EBSD (crystallography) and EDS (chemistry). The schreibersite precipitates were found to have an average size of \SI{27}{\micro m}, with precipitate sizes ranging from 15 to \SI{45}{\micro m}. EDS results were as expected for kamacite and schreibersite. The kamacite matrix was composed of 88.8-\SI{94.0}{wt\%Fe} and 5.8-\SI{6.0}{wt\%Ni}, any remainder was attributed to be carbon. The schreibersite precipitates had elemental composition of 42.6-\SI{45.2}{wt\%Fe}, 35.0-\SI{36.5}{wt\%Ni}, 13.6-\SI{12.3}{wt\%P}, and 8.8-\SI{5.9}{wt\%C}. The map sum elemental composition of the meteorite fragment was \SI{88.2}{wt\%Fe}, \SI{5.9}{wt\%Ni}, \SI{5.8}{wt\%C}, \SI{0.1}{wt\%P}. The carbon map is not provided in Fig. \ref{Fig:BigROI} as it was found dispersed across the sample and may be from additional surface contamination.

{}
\par
\begin{figure*}[!hbt]
    \centering
    \includegraphics[width=\textwidth]{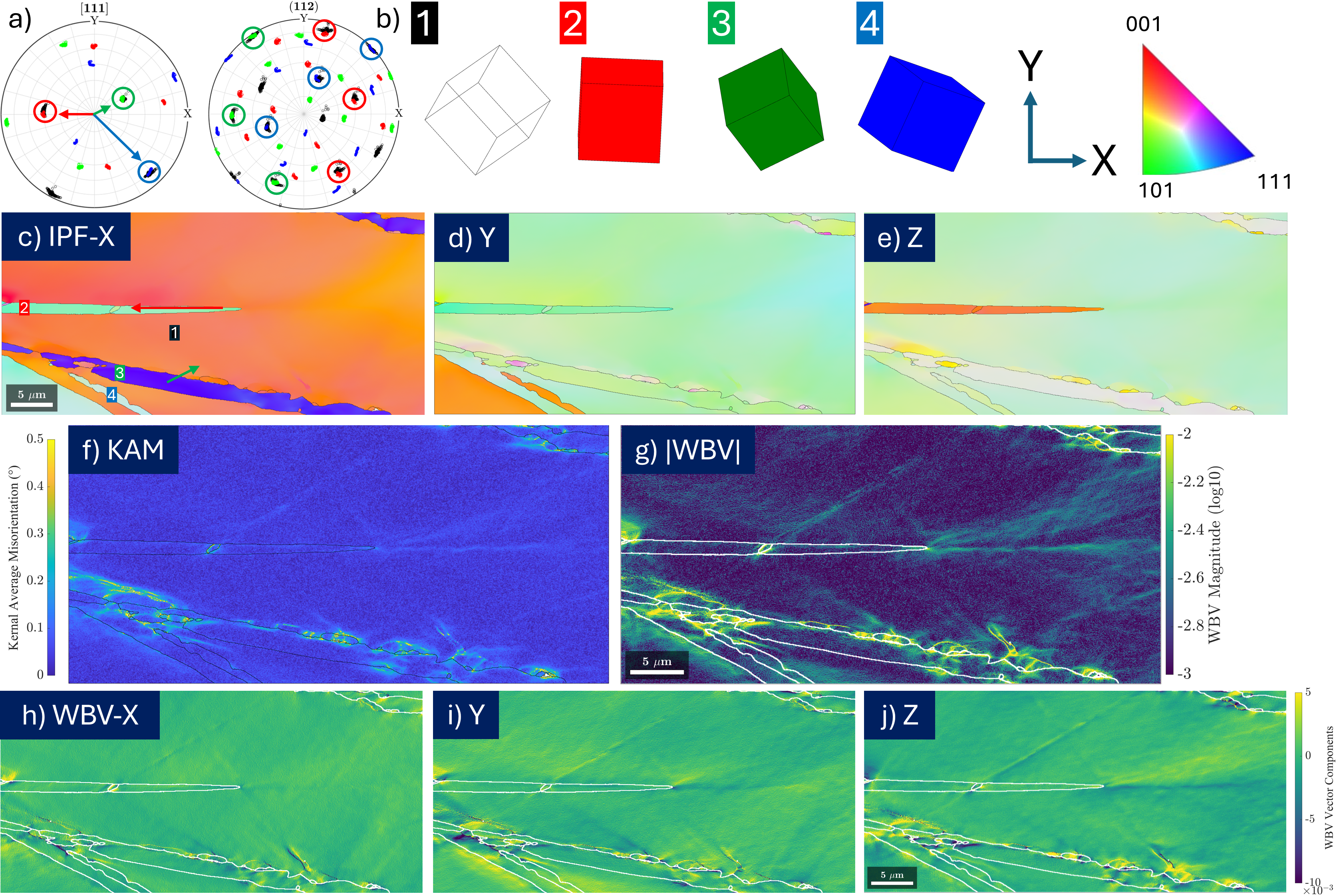}
    \caption{1st row: a) [111] and \{112\} pole figures and b) unit cells of matrix and Neumann band twins. 2nd row: IPF (c-e). 3rd row: f) Kernel Average Misorientation (KAM) and g) Weighted Burgers Vector (WBV) magnitude of isolated twin tip region with crossing twin region below. 4th row: WBV vector components (h-j). Stress regions of the isolated twin tip are visible in the IPF-X, while crossing twins high stress regions are visible in IPF-Y and IPF-Z. An increased WBV magnitude was seen around the twin grain boundaries in the lower twin. A dislocation network was seen to extend outwards from the isolated twin tip, and this is most evident in the map that reveals the WBV components as resolved with respect to the Z axis (WBV-Z).}
    \label{Fig:IsolatedTwinTip}
\end{figure*}

 Neumann bands were seen to be interlaced over the entire surface with crystal orientations being consistent amongst parallel bands. The boundaries of these bands were found to be <111> twins with an average \SI{59}{\degree} misorientation. Areas of elevated lattice rotation gradient (i.e. strain gradient) can be seen around the intersection point of the Neumann bands, shown by the slight changes of orientation in the IPF plots. These regions are most visible in the IPF-X of Fig. \ref{Fig:BigROI} and further investigation was performed at the isolated twin tip region in Fig. \ref{Fig:IsolatedTwinTip}.

{}

\par  A greater magnitude of misorientation was seen in the lower twin present in Fig. \ref{Fig:IsolatedTwinTip}. The IPF maps showed minimal orientation changes in the kamacite surrounding the isolated twin tip, however an increase in misorientation is seen extending outward from the twin tip. The kernel average misorientation (KAM) was seen to be largest in this region of the tip, though the magnitude was low (approx. \SI{0.15}{\degree}).
\begin{figure*}[!hbt]
    \centering
    \includegraphics[width=\textwidth]{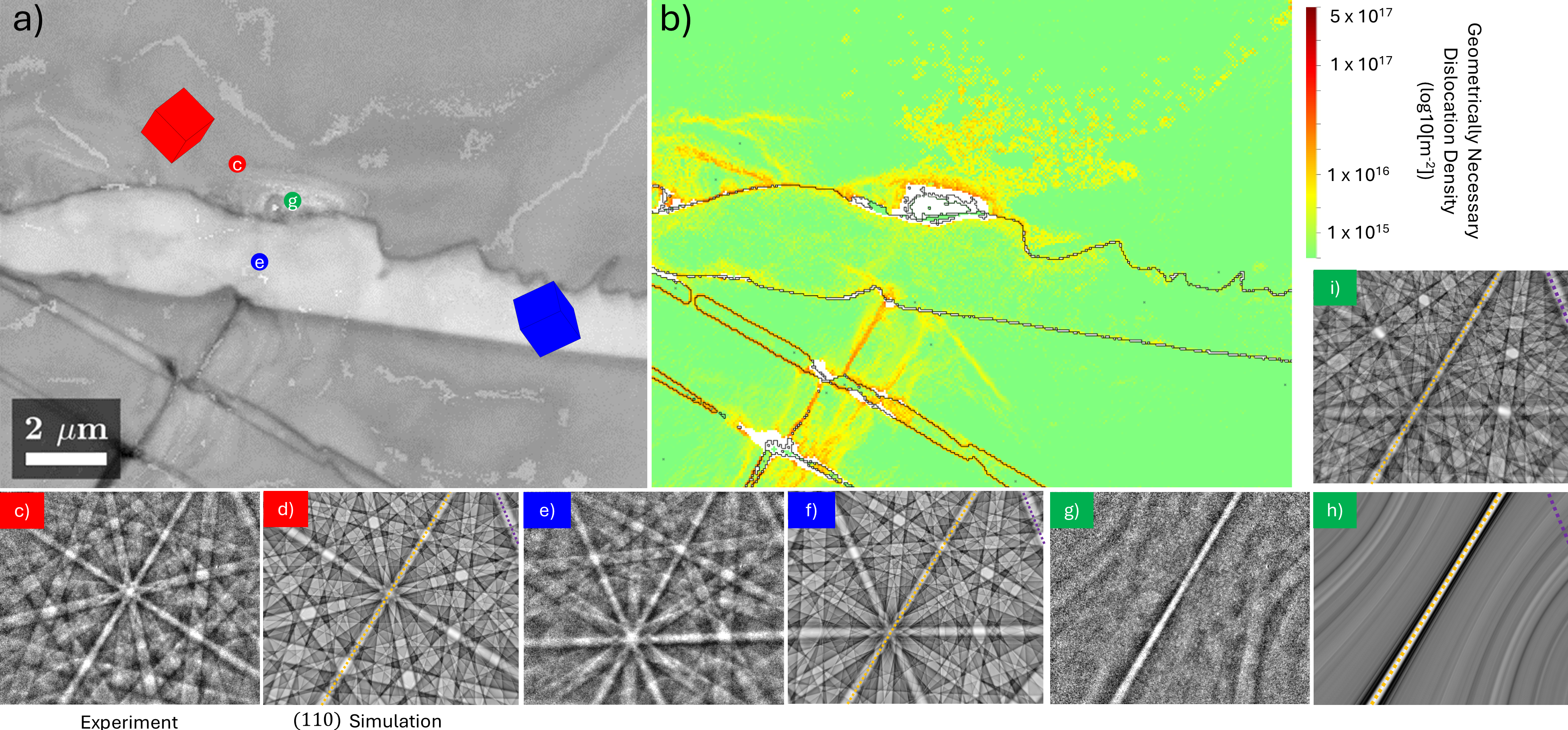}
    \caption{a) Band contrast image obtained using EBSD and b) Geometrically Necessary Dislocation Density map using HR-EBSD of high stress intersecting twin region with Electron Backscatter Patterns (EBSPs) presented from before and after the intersection. Typical BCC iron patterns are seen at (c) and (e) with the simulated EBSP to their right (d and f respectively), and the \{110\} band highlighted. Band doubling was seen at the exit point of the Neumann band intersection (g) and was seen to have a single high intensity \{110\} band with blurry lower intensity bands mirrored across it. The simulated pattern for confined and localized crystal rotation (h) and multi-layer twinning (i) are provided.}
    \label{Fig:SmallROI}
\end{figure*}{}
\par The isolated twin was found to have increased misorientation at a second location near its middle. The source of misorientation near the middle appears to be the initiation of a new twin which had not fully formed into a Neumann band. However, the embryo of a Neumann band can be seen through the KAM in Fig. \ref{Fig:IsolatedTwinTip}. The presence of these embryos is common on the sample surface and are likely indicative of the direction which a new Neumann band would form with additional shock.

\par Further analysis was performed using Weighted Burgers Vector (WBV) \cite{Wheeler2009} to analyze the dislocations present within the isolated tip ROI (Fig. \ref{Fig:IsolatedTwinTip}). The WBV magnitude reveals a greater number of dislocations within the tip front region that fan out and upward toward the twin band present in the upper right. A large WBV magnitude is observed found near the middle of the twin, of course this is also revealed as regions of measured KAM (which is related to the GND density and the WBV analysis). The presence of the potential Neumann band formation embryo was similarly visible using WBV magnitude compared to the KAM. These dislocation structures are most evident within the WBV-Z maps. 

\par Furthermore, the component maps of the WBV (i.e. WBV-X, WBV-Y, WBV-Z) show the polarization of the dislocation field associated with each twin variant. For example, the large horizontal twins (twin 2) shows that the WBV is dominated by a vector that points in Y and Z (i.e. has limited X component). The two fields that extend away from this twin, one horizontal and associated with the lower twin-matrix interface and the second running diagonally, are of opposite sign, which is important to ensure that there is closure of the deformation field further away from the twin termination site. The other twin variants show a similar effect that starts at the termination of the twin and extend into the matrix, but the WBV vector is, of course, different (as related to the twin-matrix misorientation for each variant).

\par A further region of interest was identified near the isolated twin tip, shown in Fig. \ref{Fig:SmallROI}. The region in question was identified to have a high KAM value and a large number of dislocations within a small twin intersection. Initial indexing of this location failed and direct investigation resulted in the analysis of Electron Backscatter Patterns (EBSPs) at the investigated location seen in Fig. \ref{Fig:SmallROI}.

\par Sum patterns of the BCC-iron EBSP were simulated to evaluate the potential of multilayer twinning at the twin-twin interaction point. The individual common/invariant plane \{110\} normal vector was calculated using the indexed patterns (obtained from pattern matching with the matrix pattern). A two pattern sum was generated through rotation of the matrix orientation by \SI{-70}{\degree} about this normal vector, resulting in a second pattern that was similar in orientation to the twin orientation but with an exact overlap of the invariant plane (as opposed to a propagation of uncertainty from using pattern matching of both the parent and twin experimental patterns). The original simulation and this twin simulation were summed together and the intensities normalized, resulting in the pattern given in Fig. \ref{Fig:SmallROI} i). Simulation and evaluation of possible constrained lattice rotation was performed in a similar manner. For the constrained lattice rotation, a simulation of the reference pattern was rotated about the same normal vector 360 times in \SI{1}{\degree} increments, and the sum pattern of this rotation series was calculated and the intensities normalized. The resulting pattern is provided in Fig. \ref{Fig:SmallROI} h).

\par
\begin{figure*}[!htb]
    \centering
    \includegraphics[width=\textwidth]{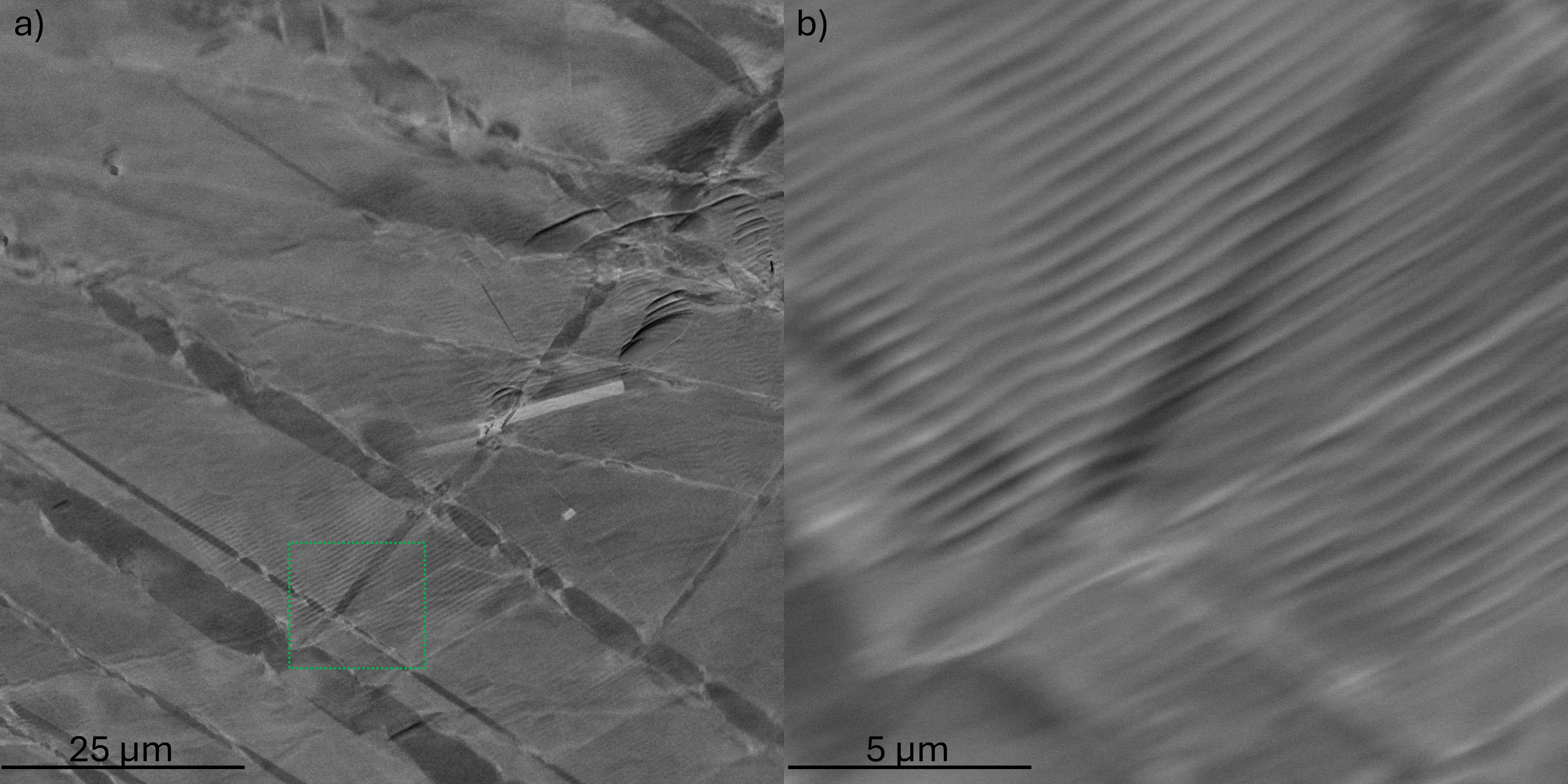}
    \caption{Forescatter electron (FSE) images of a Campo del Cielo meteorite fragment showing Neumann bands and overlapping `ripples', likely due to formation of the `ripples' after the Neumann bands. a) The ripples are seen to extend over large areas and over multiple Neumann bands; b) higher magnification FSE imaging of the `ripples'. The ripples are seen to have minor shifts, likely due to shear deformation from Neumann band growth. }
    \label{Fig:Ripples}
\end{figure*}

 FSE imaging also led to the discovery of a seemingly `rippled' surface as seen in Fig. \ref{Fig:Ripples}. While these `ripples' appear to be deformations on the material surface, they are instead surface contrast resulting from electron channelling effects through the use of FSE imaging. Further discussion of this channelling behaviour is provided in \cite{Britton2018, Winkelmann2017, Prior1996}. These `ripples' were seen to run parallel to one another, with a consistent spacing of approximately \SI{0.5}{\micro m}. The ripples were found to be present over large distances and were shifted when near Neumann bands.

\begin{figure*}[!htb]
    \centering
    \includegraphics[width=\textwidth]{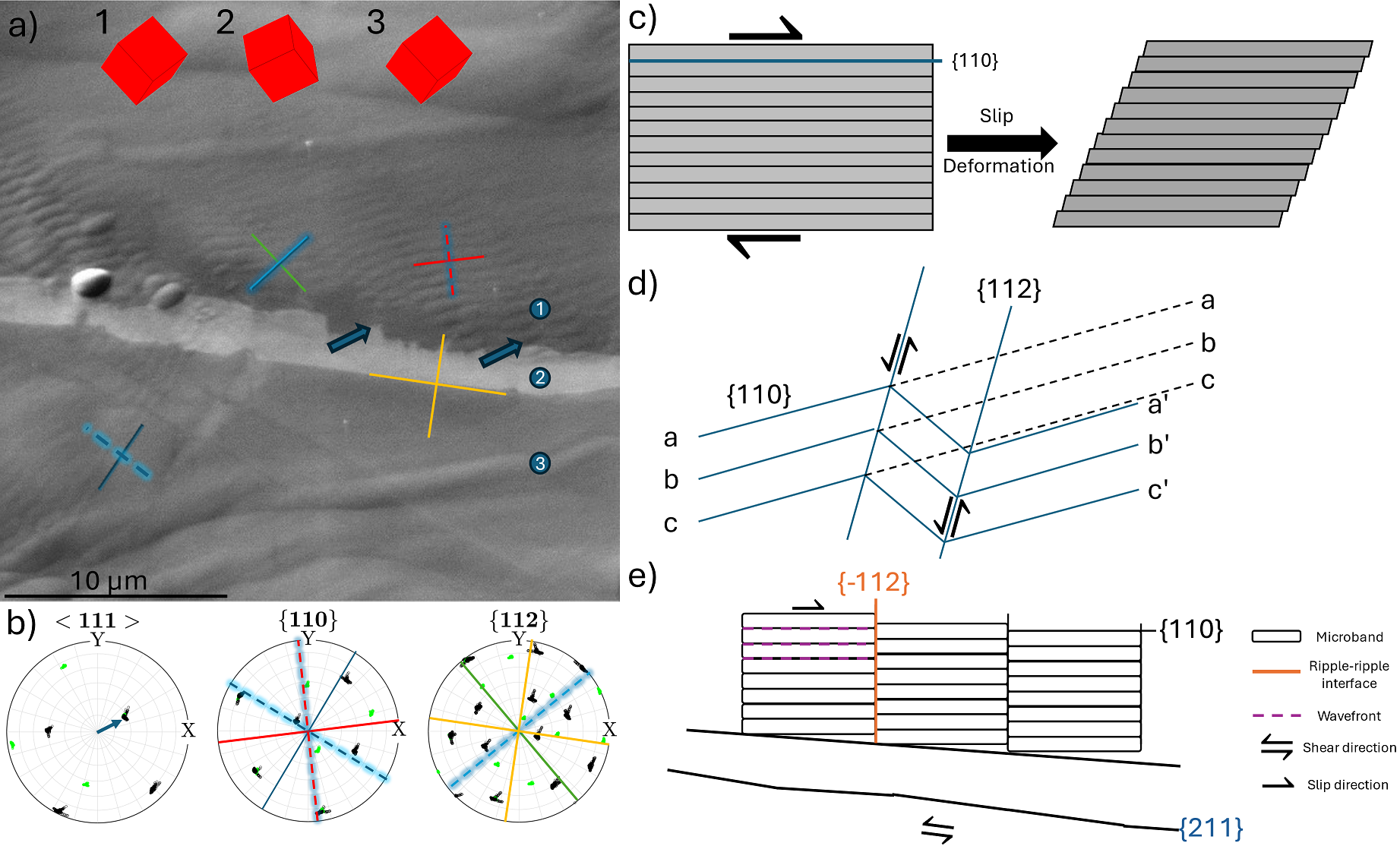}
    \caption{a) SEM micrograph of the meteorite surface showing Neumann Band like ripples with unit cells of the kamacite (BCC iron) matrix and Neumann band. b) Pole figures for <111>, \{110\} and \{112\} are presented with surface trances relating to the ripple pattern overlaid on a). The red and blue traces across the `wavefront' of the ripples are found to align with the \{110\} plane. The green-blue trace is aligned with the ripple-ripple interface and overlaps with the \{112\} plane. The ripples are seen to form interfaces similar to the planes in the Neumann band mechanism. The `wavefront' of the ripples are found to match the slip plane of BCC. The general formation mechanisms for slip bands and Neumann bands are given in c) and d), respectively. For c), the slip bands form along the \{110\} plane. In d), the Neumann bands form through shear deformation along the \{112\}, with a, b, and c being a \{110\} plane. The resulting structure from both mechanisms is provided in e).}
    \label{Fig:RipplesMechanism}
\end{figure*}

A high magnification analysis of the `ripples' was performed and is provided in Fig. \ref{Fig:RipplesMechanism}. The `ripples' were seen to extend parallel to Neumann bands present in the material surface but did not appear to change direction when crossing large Neumann bands. Pole figures were created for the \{110\}, and \{112\} planes and [111] direction. Surface traces indicated that the `wavefront' (red-blue and blue-blue traces, Fig. \ref{Fig:RipplesMechanism}) of these ripples aligned with \{110\}. The ripple-ripple interface was found to align with \{112\} (green-blue trace, Fig. \ref{Fig:RipplesMechanism}). 

\section{Discussion}

\par This Campo del Cielo fragment was found to be covered in Neumann bands, with numerous intersections points between these bands. The measured thickness of the Neumann bands varied between \SI{0.2}{\micro m} and \SI{18}{\micro m}, with an average misorientation angle of \SI{59}{\degree}. Closer investigation of isolated Neumann bands/twins revealed a network of dislocations (Fig. \ref{Fig:IsolatedTwinTip}) from an isolated twin tip to the Neumann band twin present above it and a higher WBV magnitude, and therefore dislocation density, along the boundary of the lower Neumann band. 

\par The dislocation density on the lower Neumann band varied along the boundary, with areas of high density dotted along the upper side of the twin grain boundary. It is noted that this increased density is not observed in the IPF-Y and Z (Fig. \ref{Fig:IsolatedTwinTip}) on the lower side of the grain boundary. 

\par This notched appearance on the upper twin boundary has previously been observed by \emph{Mahajan and Bartlett} \cite{Mahajan1971} for shock deformation twins in molybdenum; however, the cause was undetermined. A mechanism for nanoscale BCC tantalum has been proposed by \emph{Zhong et al.} which explained a similar difference in the twin boundaries through competing nucleation-controlled and growth-controlled mechanisms \cite{Zhong2024}, but these mechanisms have not been validated for the micrometre and millimetre scales of the Neumann bands. 

\par Rather, the Neumann bands likely initially formed through the slip dislocation interaction mechanism \cite{Priestner1965, Li2023b}, as observed in Fe-\SI{2}{\%Ni} and Fe-\SI{4}{\%Ni} alloys previously \cite{Priestner1965}. It is noted that no intersecting slip bands were found during microstructural analysis of the Campo del Cielo fragment. This further corroborates that the presence of slip bands is not required for deformation twin formation \cite{Mahajan1971, Mahajan1971b}.  

\par The notched appearance of the band may be a result of discrete stepwise growth of the twin boundary along \{112\} in a similar fashion to the mechanism described by \emph{Jiang et al.} \cite{Jiang2018}. Alternatively, the notched appearance of the twin boundary may be due to the concurrent formation of many smaller deformation twins which subsequently grew together, as seen in high strain rate deformation of bulk-polycrystalline BCC tantalum \cite{Chen2014}.   

\par The isolated twin tip with the network of dislocations is not seen to have this notched appearance. Rather, the twin tip has a relatively flat boundary with a lower KAM and WBV magnitude along its body, and increased magnitudes at and near the tip. It is possible that the increased dislocation density near the tip is a precursor to new grain boundary formation through a nucleation mechanism. Similar behaviour has been simulated previously by \emph{Sainath and Nagesha} for the nucleation and initiation of a BCC iron twin \cite{Sainath2022}. However, if this twin tip were to further propagate it would be outward from the tip point, rather than upward toward the Neumann band in the upper right of the WBV magnitude in Fig. \ref{Fig:IsolatedTwinTip}. 

\par Comparison of the KAM map and WBV maps in Fig. \ref{Fig:IsolatedTwinTip} are interesting as: 
\begin{itemize}
    \item[(1)] The WBV provides higher contrast to show areas of heterogenous deformation, as comparing f) and g).
    \item[(2)] The decomposed vector components (WBV-X, Y and Z – shown in h-j) reveal that dislocation density is polarized around the intersections and tips of the Neumann bands, including those bands that are not revealed as distinctly indexed features.
    \item[(3)] The polarization of these fields for each of the similar Neumann band variants is similar, which is consistent with what would be expected for conventional deformation twinning that occurs in more conventional deformation experiments (e.g. as reported in Guo \emph{et al.} \cite{Guo2017}).
\end{itemize}
\par In this work, the WBV is used to quickly analyse and describe components of the deformation gradient that can be accessed quickly from the measured maps. Overall, we have presented Fig. \ref{Fig:IsolatedTwinTip} as an opportunity to provide new insight into the deformed meteorite sample, as well as providing a comparison of the different modes of analysis that are now routinely available within modern EBSD packages.

\par Further analysis of the smallest region of interest lead to the discovery of the unusual EBSP shown in Fig. \ref{Fig:SmallROI} in this region of very large deformation near the edge of the twin-twin interface. The twin and matrix EBSPs show that there is a common  \{110\} band within these diffraction patterns, and the unindexed region (in region 3 of the map) has a pattern which contains a single diffraction band superimposed upon a noisy 'background' (which has some evidence of other diffracting features). In general, in the presence of a strain gradient within the interaction volume, it is well known and well understood that the EBSD-pattern can blur and become harder to index. In this unusual example, all but one band has blurred, and this indicates that the strain gradient for this volume is highly restricted and there is crystalline material with a (relatively) undisturbed \{110\} plane and one could imagine that this is the extreme case of 'pattern doubling' that could be related to either a very high strain gradient or multilayer twinning \cite{Li2023} in this localized region. Further work would be required to qualify the interaction volume size and the minimum number/amount of pattern blurring that is required to try to estimate the strain gradient required for the exact amount of blurring seen in the experiment, which is out of the scope of the present work. To assist in understanding the rest of the deformation in this area, this motivated the HR-EBSD based GND analysis, which supports the idea that there is a localized strain gradient in this location due to the presence of a very high GND density near this region.

\par A multilayer twin is a section of the material where multiple twin deformations are stacked atop one another. These multilayer twins are denoted by the number of packing layers between the twins, with density of twin increasing as the number is lower \cite{Li2023}. For example, a 9-layer twin has 9 layers (ABCABCABC) between it and the twin deformation plane (illustration provided in \cite{Li2023}). When the electron beam direction is parallel to the twin plane of this stack it results in overlap of EBSP of each twin, causing constructive interference near the twin plane band and increasingly destructive interference further away \cite{Li2023}. This behaviour has been simulated previously by \emph{Li et al.} for FCC iron down to 3-layer twins. However, the simulated EBSP given in Fig. \ref{Fig:SmallROI} i) was not seen to match the experimental EBSP (\ref{Fig:SmallROI} g). Rather, the rotation series pattern (\ref{Fig:SmallROI} h) looks more similar to the experimental pattern. Notably, this contains the presence of a bright \{110\} band with no features of zone axes along the band, as well as blurring of the other bands in the pattern. For the 'double' pattern shown in Fig. \ref{Fig:SmallROI} i), the nature of this orientation relationship would leave the \{110\} band in the top right of the pattern also reproduced exactly in the experiment, which was not observed.

\par From this simulation, it is believed that constraint of the crystal structure and localized crystal rotation play a part in the high \{110\} presence seen in the EBSP. Intersection of the two twin bands results in the generation of shear stress and movement through specific slip systems. Those being \{110\}, \{112\}, and \{123\} for BCC metals, in order of increasing rarity \cite{Weinberger2013}. The most common slip plane, \{110\}, is known to be related to the formation of Neumann bands, with \{110\} acting as a twin plane for the formation of Neumann bands \cite{Smith1928}. It is suggested that the twin-twin intersection and the shear stress involved results in slip along the \{110\} plane, constraining the microstructure and its dislocations. Thereby, causing a high intensity \{110\} band to be seen at the site of exit for the twin-twin intersection.

\par FSE imaging of this region also showed an interesting surface phenomena. A ripple-like appearance was seen (Fig. \ref{Fig:RipplesMechanism}) that seemed to travel along planes similar to the large Neumann bands. Surface traces on the \{110\} and \{112\} pole figures were performed, given the known relation of these planes to Neumann bands \cite{Uhlig1955, Smith1928}. This revealed an overlap with the \{110\} plane and `wavefront' of the ripple, and for the \{112\} plane with the ripple-ripple interface. Additional ripples were seen over large regions of the sample, shown in Fig. \ref{Fig:Ripples}, that have the same `wavefront', but are not seen to have the same ripple-ripple interfaces as in Fig. \ref{Fig:RipplesMechanism}. 

\par The relation seen for the `wavefront', or walls, of these `ripples' with \{110\} and their presence in and around high strain regions suggests that these may be microbands. The development of microbands in BCC-iron and other BCC materials is a known phenomena induced through rapid shock \cite{Huang1988, Thuillier1994}, which forms along BCC slip systems (\{110\}, \{112\}) \cite{Huang1988, Dorner2007}. Similar behaviour is found here with the `wavefront' coinciding with the preferential slip plane of microband formation. The ripple-ripple interface seen along \{112\} is believed to be a result of deformation twin growth, resulting in shear deformation of the surrounding matrix. An illustration is provided in Fig. \ref{Fig:RipplesMechanism} e) to show the relationship of the ripples to the slip direction, shear direction, and the microbands. 

\par This would indicate that the ripples and Neumann bands did not form simultaneously. The overlapping of these ripples with the Neumann bands in Fig. \ref{Fig:Ripples} and lack thereof in Fig. \ref{Fig:RipplesMechanism} a) further reinforces this. The lack of overlap seen in the latter is indicative that the deformation twin formed after the ripples, likely due to a different shock event. Different types of shock has been found in the past to affect what microstructural features will develop in FCC materials, with plane shock and spherical shock inducing deformation twinning and microband formation, respectively \cite{Murr2004}. While this relationship was developed using FCC materials, it has been proposed by \emph{Huang and Grey} that the formation mechanism of microbands may be similar in FCC and BCC materials \cite{Huang1989}. Rather, the specific material dictates the propensity for strain accommodation to occur through microband formation or deformation twinning. 

\section{Conclusion}

\par The Campo del Cielo meteorite fragment was found to be composed of a kamacite matrix with blocky schreibersite precipitates present throughout. Deformation twin grains, called Neumann bands, were found to cover the surface  of the sample with a high amount of intersection. These Neumann bands and their intersections resulted in high stress and strain regions within the sample surface that were detectable through the use of EBSD-based analysis.  

\par An investigation of isolated Neumann bands using WBV analysis revealed a network of dislocations extending outward from an isolated twin tip and spots of high dislocation density along one side of a Neumann band. The high dislocation density and elevated strain energies on a single twin boundary were believed to be a result of twin boundary migration through a stepwise growth mechanism or incipient twin intergrowth.  Additionally, networks of dislocations were seen to extend outward from the isolated twin tip and are believed to be indicative of further twin propagation through nucleation. 

\par Closer investigation was performed into a high strain region of the isolated Neumann bands where a twin-twin intersection was discovered to be the source. Unusual EBSPs were found at the exit point of this intersection with a high intensity \{110\} band with blurry and lower intensity bands being mirrored across the \{110\} band. It is suggested this phenomenon is a result of constrained and localized rotation of the crystal structure about the \{110\} plane due to dislocation slip associated with the Neumann band-based twin-twin interactions.  

\par Additionally, a unique surface phenomenon was seen around the Neumann bands upon high magnification imaging at multiple locations of the sample. A rippled microstructural feature was found surrounding the Neumann bands, or overlapping with them in other instances. These ripples were all seen to have a `wavefront' which coincided with the \{110\} plane of kamacite, and some with a ripple-ripple interface coinciding with the \{112\} plane of kamacite. These microstructural features are suggested to be a microband strain accommodating feature within the iron meteorite which are formed through slip deformation. Furthermore, it is suggested that the formation of the Neumann bands and the ripples occurred due to separate shock events, planar and spherical shock respectively. 

\par To summarize, EBSD-based investigation of high strain Neumann band regions in the Campo del Cielo meteorite fragment revealed two highly unusual phenomena, the `ripple' microbands across the surface and the mirrored high intensity \{110\} band EBSP present at a Neumann band intersection. This study also reports the coexistence of Neumann bands (microtwins) and microbands within an iron meteorite. These findings may be significant in furthering our understanding of high strain rate microstructural changes within both terrestrial and interstellar iron-based BCC materials under extreme shock conditions.

\section*{Acknowledgements}
    We acknowledge the support of the Natural Sciences and Engineering Research Council of Canada (NSERC) [Discovery grant: RGPIN-2022–04762, ‘Advances in Data Driven Quantitative Materials Characterisation’]. The AMBER-X pFIB-SEM was funded by the B.C. Knowledge Development Fund (BCKDF) and Canadian Foundation for Innovation-Innovation Fund (CFI-IF) (\#39798: AM+). We also acknowledge Haroon Qaiser and Tianbi Zhang for their helpful discussions in creating this work.

\section*{CRediT}

\textbf{Graeme J. Francolini:} Conceptualization, Formal Analysis, Investigation, Methodology, Writing - Original draft, Writing - review \& editing. \textbf{Thomas. B. Britton} Formal Analysis, Funding acquisition, Supervision, Writing - review \& editing.

\section*{Data Availability}
Data will be released to Zenodo when the article is accepted for publication, and a DOI will be inserted here.
\bibliographystyle{elsarticle-num}

\bibliography{article.bib}

\end{document}